\documentstyle[12pt,aps,preprint,psfig,floats]{revtex}
\begin{document}
\title{KINETIC THEORY OF THE OVERLAPPING PHASE TRANSFORMATIONS}
\author{Valencia-Morales, E.\thanks{evalen@mfc.uclv.edu.cu}, Galeano-Alvarez, N. J., Vega-Leyva, J., Villar, C, E., and Hernandez-Ruiz, J}
\address{Department of Physics, Central University of Las Villas, Santa
Clara 54830, CUBA.}
 \date{\today}
 \maketitle

\begin{abstract}

  Often a number of precipitation processes in steels and alloys occur simultaneously (they have the same origin in time) albeit at different rates. Consequently, isothermic transformations are accompanied, in this case, by the occurrence of second processes of precipitation that influence the changes of the macroscopic parameter chosen for the kinetic study while first processes are occurring. In this context a set-up is presented that addresses the issue of characterization of nucleation and growth mechanisms of the individual processes. Soft and hard interferences of processes are considered. The present approach is based on the Avrami model and allows computing the kinetic parameters (n, k) of the different precipitation processes in isothermic phase reactions.  \\
\end{abstract}.
\newpage
\section{Introduction}

  The Johnson-Mehl-Avrami(JMA) equation has been widely used to study the kinetics of the nucleation and growth of a second phase from an initial parent phase \cite{Wilkes,Murphy,Darbyshire,Lai,Saito}. This kinetic relation needs assumptions about the nucleation and growth rates of the new phase nuclei \cite{Johnson,Avrami1,Avrami2,Avrami3,Christian}. By evaluating the JMA equation, an index n is obtained. Knowledge of n, together with the empirical value of the activation energy and some other features of the transformation, yields a much better understanding of the reaction mechanism.

  In practice, a macroscopic parameter P(t) is commonly used for the kinetic study of the
  isothermic reactions \cite{Christian}. By simplicity, it is assumed that P(t) depends
  linearly with the solute concentration c(t) that participates in the reaction. Variations of P(t) during the transformation are only associated with a type of change; either by entry or exit of the solute in the parent phase or by formation /dissolution of a second phase. The phase fraction that is transformed in time can then be written as: \\

\begin{equation}
\xi(t)=\frac{\triangle P(t)}{\triangle P(t_{end})}.
\end{equation}

  Often, in practical situations as the low-alloy steels martensite tempering and many other isothermal phase transformations in steels, precipitation processes do not occur isolately \cite{Bang,Bhadeshia,Bhadeshia1}. Instead, first and second processes develop simultaneously (the same origin in time) albeit at different rates. Then, the macroscopic experimental parameter P(t) contains information of both processes (overlapping processes). In this sense, some authors \cite{Prado}\ have remarked evident difficulties while trying to set apart the effects of each one of these processes by knowledge of only the macroscopic parameter P(t).

  Most of these works described above have involved fitting of an Avrami type equation to the
  transformations curves for a specific alloy at specific temperatures. These have considered
  only one precipitation reaction neglecting the influence of the other precipitation processes
   from the experimental data. Therefore, the variation of the fraction transformed over time
   is frequently found to deviate from a simple sigmoidal relationship( like JMA behavior), because the curve    $\xi(t)$ vs. $\ln{t}$ actually represents more than one reaction occurring at the same time.

There have been a number of attempts to deconvolute such master
curves into components due to individual processes
\cite{Hanawaka,Luiggi,Luiggi1,Luiggi2}, but it is difficult to see
whether some of these fitting parameters have physical meaning.

The aim of the present paper is, precisely, to outline a set-up based on the Avrami model, that
 allows obtaining the kinetic parameters (n, k) for the simultaneous precipitation processes
  with interference by the only knowledge of a macroscopic  parameter P(t) chosen for the
  kinetic study. This new set-up will be helpful for the interpretation of the experimental
  data. It does not fall into the category of the predictive models of references \cite{Robson,Jones,Fujita}.\\

\section{Kinetic Theory of Two Overlapping Processes.}

  Assuming that JMA equation holds for individual processes is equivalent to considering only hard impingement between particles that participate in each process. In other words, the soft impingement for a given type of particles in the same spatial region is not considered.

Let us take two processes that rival for the same solute at different rates. Suppose both obey JMA equations. Then, the real fraction transformed that is determined by the changes of the experimental parameter obeys a relation that is similar to Eq. (1):\\

\begin{equation}
\xi_{r}(t)=\frac{\triangle P(t)}{\triangle P(t_{end1})},
\end{equation}
\\where $t_{end1}$ is the time at the end of the first process. In Eq. (2) we chose $t_{end1}$ instead of $t_{end2}$  because we take into account the effect of the first process during the interval $t\leq t_{end1}$ when there is appreciable influence of the second process on the experimental parameter. When the existence of a second process is known, it is used to fix the end of the reaction at the time when the process that causes major influence in the parameter P(t) ends. In the case when this last is the second process, it is used to neglect variations in P(t) due to the first one. Obviously it is incorrect.

It is interesting finding the relation between $\xi_{r}(t)$ and
the fractions $\xi_{1}(t)$   and $\xi_{2}(t)$ of the elementary
processes by knowledge of $\triangle P(t)$ only.

Let us consider that variations of P(t) associated with each process are independent, i.e.,\\

 \begin{equation}
\triangle P(t)=\triangle P_{1}(t)+\triangle P(t_{2}).
\end{equation}

Therefore, we neglect the reciprocal interference of both
processes, i.e., the overlapping of the drainage zones of the
chemically different precipitates that correspond to processes 1
and 2. The bigger the differences of the rates at which the
individual processes take place, the higher the reliability of
this approximation. Neglecting the correlation effects in Eq.(3)
is less significant than neglecting soft impingement for
precipitates of a same kind since, phase transformations in both
processes, often, take place at different geometrical places. In
case that the rate of one of the processes is close to that of the
other process, rivalry for the same solute leads to the situation
when a phase alone rivals for the same solute, in other words, a
situation with soft impingement among particles of a given kind.
But we recall that the present set-up, soft impingement are
neglected, i.e., JMA-type expressions are satisfied.

Therefore, for each elementary process we have:\\

\begin{equation}
\xi_{1}(t)=\frac{\triangle P_{1}(t)}{\triangle P_{1}(t_{end1})},
\end{equation}
\\and\\

\begin{equation}
\xi_{2}(t)=\frac{\triangle P_{2}(t)}{\triangle P_{2}(t_{end2})},
\end{equation}
\\respectively so,\\

\begin{equation}
\xi_{r}(t)=\frac{\triangle P_{1}(t)+\triangle P_{2}(t)}{\triangle
P_{1}(t_{end1})+\triangle P_{2}(t_{end1})}.
\end{equation}

If we take into account Eqs.(4) and (5), then Eq.(6) can be written as,\\

\begin{equation}
\xi_{r}(t)=\frac{\xi_{1}(t)+\alpha_{p}\xi_{2}(t)}{1+\alpha_{p}\xi_{2}(t_{end1})},
\end{equation}
\\where\\

\begin{equation}
\alpha_{p}=\frac{\triangle P_{2}(t_{end2})}{\triangle
P_{1}(t_{end1})}.
\end{equation}

The magnitude $\xi_{2}(t_{end1})$ measures the degree of
overlapping (soft or hard) of both processes. Depending on the
choice of P(t), the parameter $\alpha_{p}$ may be either negative
or positive. When P(t) moves in contrary senses for both
processes, $\alpha_{p}<0$ but when P(t) moves in the same sense,
$\alpha_{p}>0$.

We remark that $\xi_{r}(t)$ depends on the parameter chosen for the kinetic treatment of the processes. Recall that $\alpha_{p}$ takes different values for the different macroscopic parameters P(t) one chooses.\\

\subsection{Computation of $\xi_{r}(t)$ with the help of the isothermic dilatometric records. }

Dilatometric records are often used for the characterization of
the transformation kinetics of a significant group of steels
\cite{Darbyshire,Roberts,Takahashi,Dirand}. The experimental
determination of $\xi_{r}(t)$ would be possible if we would be
able to measure the parameter P(t)=l(t) from the beginning of the
phase transformations (in an isothermic regime). However, the
sample needs some time for reaching the temperature of the
isothermic treatment (for example, the temperature of isothermic
tempering of martensite structures). During this small time
interval, the sample suffers a thermal treatment, so the beginning
of the transformations may be prior to that of the temperature
stabilization corresponding to the isothermic treatment and
therefore, the beginning of the precipitation precesses is
unknown.

Without loss of generality, the figure 1 shows a scheme corresponding to a dilatometric record belonging to a low alloy steel martensite tempering. In the interval $0\leq t \leq t_{0}$ (fig. 1), a dilatation of the sample occurs that is originated by the change in the temperature from a room temperature up to the temperature of the isothermic treatment. Simultaneously, a contraction of the sample takes place in this interval that is a consequence of a first process in the chain of transformations during the heating of the sample. Consequently, it is necessary to begin the study not from the origin of the record given by the measuring equipment but from the moment of time when the isothermic regime is reached. It is given by $l^{'}(0)$ in fig.1. If we take $l^{'}(0)$ to be the initial length of the sample ($l^{'}(0)= l(t_{0})$)then we may compute the fraction transformed through:\\

\begin{equation}
\xi^{'}(t^{'})=\frac{l^{'}(t^{'})-l^{'}(0)}{l^{'}(t^{'}_{end1})-l^{'}(0)}.
\end{equation}

We remark that $\xi^{'}(t^{'})$ is not the real fraction transformed $\xi_{r}(t)$. This last magnitude may be written, following Eq.(2) with P(t)=l(t) as:\\

\begin{equation}
\xi_{r}(t)=\frac{l(t)-l(0)}{l(t_{end1})-l(0)}.
\end{equation}

Since $\xi_{r}(t)$(Eq.(10)) can not be obtained directly from
experiment (we do not known the origin (in time) of the
transformations), it is necessary to find a relation between the
fraction transformed $\xi^{'}(t^{'})$, that is experimentally
measurable, and $\xi_{r}(t)$ . It can be shown that:

\begin{equation}
\xi(t)=\xi^{'}(t^{'})=\frac{\xi_{r}(t)}{1-\xi_{r}(t_{0})}-\frac{\xi_{r}(t_{0})}{1-\xi_{r}(t_{0})},
\end{equation}
\\where:\\

\begin{equation}
\xi^{'}(t^{'})=\xi^{'}(t-t_{0})=\xi(t).
\end{equation}

From Eq.(11) we obtain:\\

\begin{equation}
\frac{d\xi^{'}(t^{'})}{d\ln
t^{'}}=\frac{d\xi^{'}_{r}(t^{'})}{(1-\xi_{r}(t_{0}))d\ln t^{'}},
\end{equation}
\\since:\\

\begin{equation}
\xi_{r}(t)=\xi_{r}(t^{'}+t_{0})=\xi_{r}^{'}(t^{'}).
\end{equation}

If we take the derivative of Eq.(7) and considering that $\xi_{r}(t)=\xi^{'}_{r}(t^{'})$ and $\xi^{'}(t^{'})=\xi(t)$ the following expression is obtained:\\

\begin{equation}
\frac{d\xi^{'}_{r}(t^{'})}{d\ln
t^{'}}=\frac{1}{1+\alpha_{l}\xi^{'}_{2}(t^{'}_{end1})}[\frac{d\xi^{'}_{1}(t^{'})}{d\ln
t^{'}}+ \alpha_{l}\frac{d\xi^{'}_{2}(t^{'})}{d\ln t^{'}}],
\end{equation}
\\but\\

\begin{equation}
\frac{d\xi^{'}_{i}(t^{'})}{d\ln
t^{'}}=\frac{t^{'}}{t^{'}+t_{0}}\frac{d\xi_{i}(t)}{d\ln t}
    \mbox{,(i=1,2) }.
\end{equation}

Let $\xi(t)$ be the fraction transformed of an elementary process obeying a JMA expression, $\xi(t)=1-exp-(k t)^{n}$, then:\\

\begin{equation}
\frac{d(ln ln\frac{1}{1-\xi})}{d\xi}\frac{d\xi}{d\ln t}=n
\:\:\mbox{or} \:\:\frac{d\xi(t)}{d\ln t}=n Z(\xi),
\end{equation}
\\where\\

\begin{equation}
Z(\xi)= [\frac{d(ln ln\frac{1}{1-\xi})}{d\xi}]^{-1}=
-(1-\xi)ln(1-\xi).
\end{equation}

Therefore, such as $\xi_{1}(t)$ and $\xi_{2}(t)$ both obey JMA-type expressions, we obtain:\\

\begin{equation}
\frac{d\xi_{i}(t)}{d\ln t}=n_{i} Z(\xi_{i})   \:\:\mbox{,(i=1,2)},
\end{equation}
\\yielding that\\

\begin{equation}
\frac{d\xi^{'}_{i}(t^{'})}{d\ln t^{'}}=\frac{t^{'}}{t^{'}+t_{0}}
n_{i} Z(\xi_{i}),
\end{equation}
\\where the condition $\xi^{'}(t^{'})= \xi(t)$ has been taken into account. Substitution of Eqs.(20) and (15) into Eq.(13) yields a general expression correlating $\xi^{'}(t^{'})$ with the fractions of the individual processes that obey JMA-type equations:\\

\begin{equation}
(1-\xi_{r}(t_{0}))(1+\alpha_{l}\xi^{'}_{2}(t^{'}_{end1}))\frac{d\xi^{'}(t^{'})}{d\ln
t^{'}}=\frac{t^{'}}{t^{'}+t_{0}}[n_{1}
Z(\xi_{1})+\alpha_{l}n_{2}Z(\xi_{2})].
\end{equation}

The behavior of the function $Z(\xi)$ vs. $\xi$ is shown in fig. 2. This function is not symmetric respecting its maximum at $\xi=0.632$. It grows when $\xi$ increases and when $\xi\rightarrow 1$ it decreases very rapidly. However $Z(\xi)$ is nearly constant for $\xi$ lying in the neighborhood of the point where this function is a maximum. In other words, we have an interval where $\xi$ takes values (far from the boundary points $\xi=0$ and $\xi=1$) for which   $Z(\xi)$ depends weakly on $\xi$. This is a focal point of our set-up. If we allow time intervals where $\xi_{1}(t)$ and $\xi_{2}(t)$ lie far enough from the boundary points $\xi=0$  and $\xi=1$, then we may consider that $Z(\xi_{1})$ and $Z(\xi_{2})$ are nearly constants. The former requirement implies, also, that $t_{0}<<t^{'}$ so it may be neglected when compared with  $t^{'}$. Consequently Eq.(21) may be simplified:\\

\begin{equation}
(1-\xi_{r}(t_{0}))(1+\alpha_{l}\xi^{'}_{2}(t^{'}_{end1}))\frac{d\xi^{'}(t^{'})}{d\ln
t^{'}}=n_{1} Z(\xi_{1})+\alpha_{l}n_{2}Z(\xi_{2}).
\end{equation}

Two cases are of particular interest and we shall analyze them in what follows.\\

\subsubsection {Soft overlapping of the processes:}

A second process occurs, but it manifests itself only weakly during the interval
 $t_{0}<t<t_{end1}\:\: (0<t^{'}<t^{'}_{end1})$, i.e., $\xi_{2}(t_{end1})$ is small so,
 the second process perturbs the first one. In this case, since $\xi_{2}(t_{end1})$ is small, we may assume that $Z(\xi_{2})\approx 0$ for $t_{0}<t<t_{end1}$.\\
Therefore, Eq. (22) can be written as:  \\

\begin{equation}
(1-\xi_{r}(t_{0}))(1+\alpha_{l}\xi^{'}_{2}(t^{'}_{end1}))\frac{d\xi^{'}(t^{'})}{d\ln
t^{'}}=n_{1} Z(\xi_{1}),
\end{equation}
\\or\\

\begin{equation}
\frac{d\xi^{'}(t^{'})}{d\ln t^{'}}=nZ(\xi_{1}),
\:\:\mbox{with}\:\:
n=\frac{n_{1}}{(1-\xi_{r}(t_{0}))(1+\alpha_{l}\xi^{'}_{2}(t^{'}_{end1}))}.
\end{equation}

In this case when we have a weak perturbation of the macroscopic
parameter by a second process in the interval  $ t_{0}<t_{a}\leq
t\leq t_{b}<t_{end1}$ (or $0<t^{'}_{a}\leq t^{'}\leq
t^{'}_{b}<t^{'}_{end1}$), where $ t_{a}$ and $ t_{b}$ (or
$t^{'}_{a}$ and $t^{'}_{b}$)are far from the boundary points,
$\xi(t)$ (or $\xi^{'}(t^{'})$) is always close to $\xi_{1}(t)$
and, consequently, $Z(\xi_{1})\approx Z(\xi^{'})$ without taking
into account the stability region. Therefore,
the following equation takes place (see Eq.(24)):\\

\begin{equation}
\frac{d\xi^{'}(t^{'})}{d\ln t^{'}}=n Z(\xi^{'}).
\end{equation}

In this time interval there exists a linear dependence of
$lnln\frac{1}{1-\xi^{'}} \:\:\mbox{vs}\:\:ln t^{'}$ so, the slope
of this dependence is, precisely, n. In other words, in the time
interval where the best  fit of the $lnln\frac{1}{1-\xi^{'}}
\:\:\mbox{vs}\:\:ln t^{'}$ is manifest, the experimental data
satisfies a nearly JMA behavior.

Analogously, for times $t_{end1}< t <t_{end2}$, $(t^{'}_{end1}<t^{'}<t^{'}_{end2})$, when the
first process end up ($\xi_{1}=1 \:\: \mbox{and}\:\:Z(\xi_{1})=0$) we have:\\

\begin{equation}
(1-\xi_{r}(t_{0}))(1+\alpha_{l}\xi^{'}_{2}(t^{'}_{end1}))\frac{d\xi^{'}(t^{'})}{d\ln
t^{'}}=\alpha_{l}n_{2}Z(\xi_{2}),
\end{equation}
\\or\\

\begin{equation}
\frac{d\xi^{'}(t^{'})}{d\ln t^{'}}=\overline{n}Z(\xi_{2}),
\:\:\mbox{with}\:\:
\overline{n}=\frac{\alpha_{l}n_{2}}{(1-\xi_{r}(t_{0}))(1+\alpha_{l}\xi^{'}_{2}(t^{'}_{end1}))}.
\end{equation}

Since $Z(\xi_{1})=0$, the second process develops alone in this
time interval, therefore $ Z(\xi_{2}) \simeq Z(\xi^{'})$, where
$\xi^{'}(t^{'})$ decreases while $\xi_{2}(t)$ increases for the
studied case ($ \alpha_{l}<0$). Then, for a certain time interval
in the neighborhood of the maximum of  $Z(\xi)$ the points
$(Z(\xi^{'})$, $\xi^{'})$ and $(Z(\xi_{2})$, $\xi_{2})$ are
symmetrically located respect to this maximum. This assumption
enables one finding a time interval where $lnln\frac{1}{1-\xi^{'}}
\:\:\mbox{vs}\:\: lnt^{'}$ is nearly linear. This means that the
experimental fraction transformed follows a nearly JMA behavior.
Eq.(27) can then be
written as,\\

\begin{equation}
\frac{d\xi^{'}(t^{'})}{d\ln t^{'}}=\overline{n}Z(\xi^{'}),
\end{equation}
\\and picking a linear arrangement allows computing $\overline{n}$ by finding the slope of this curve.

Relationships (23), (24), (26) and (27) yield a relationship between the $\alpha_{l}$  parameter, the JMA indexes for each individual process and the values $\mbox{n}$ and $\overline{n}$ found from linear adjustment of the experimental data:\\

\begin{equation}
\alpha_{l}=\frac{n_{1}\overline{n}}{n_{2}n}.
\end{equation}

\subsubsection {Hard overlapping of the processes}

In this case a second process is strongly manifested during the
time when the fist one is occurring. We start from Eq.(22) for the
time interval $t_{0} <t< t_{end1}$. As already remarked, it is
necessary to pick a time interval such that $\xi_{2}>0.25$, i.e.,
$Z(\xi_{2})$ is nearly constant and, consequently, $\xi_{1}(t)$
and $\xi^{'}(t^{'})$ are below 0.9.This means taking $t_{0}<
t_{\alpha}\leq t$ from the left and
  $t\leq t_{\beta}< t_{end1}$ from the right. Therefore for times in the interval
  $[t_{\alpha},t_{\beta}]$$[or \;t^{'}_{\alpha},t^{'}_{\beta}]$, $Z(\xi_{1})\approx\:Z(\xi_{2})\approx\:Z(\xi^{'})\approx\:\mbox{const}$ and the Eq.(22) simplifies to:\\

\begin{equation}
\frac{d\xi^{'}(t^{'})}{d\ln t^{'}}=nZ(\xi^{'}),
\:\:\mbox{with}\:\:
n=\frac{n_{1}+\alpha_{l}n_{2}}{(1-\xi_{r}(t_{0}))(1+\alpha_{l}\xi^{'}_{2}(t^{'}_{end1}))},
\end{equation}
\\Yielding that, in this time interval, the experimental data gets
linearized, i.e., $lnln\frac{1}{1-\xi^{'}} \:\:\mbox{vs}\:\:\:\:\:
lnt^{'}$ is nearly linear. The case when the first process ends up
($t_{end1} <t< t_{end2}$), is just the case of soft overlapping
already studied (Eqs.(26) and (27)).
    The relationship between $\alpha_{l}$, the indexes for individual processes
 ($n_{1}$ and  $n_{2}$) and those obtained from the linear adjustment of the experimental
data $(n,\overline{n})$ for this case of overlapping can be
established with the help
of Eqs. (27) and (30):\\

\begin{equation}
\alpha_{l}=\frac{n_{1}\overline{n}}{n_{2}(n-\overline{n})}.
\end{equation}

\subsection {Testing the kinetics of the individual processes}

Testing the kinetics of a second process allows one obtaining
$\xi_{2}(t)$ or $\xi^{'}_{2}(t^{'})$ and therefore the parameters
$n_{2}$ and $k_{2}$ can be computed with the help of the Avrami
model.

Therefore, for $t> t_{end1}(t^{'}>t^{'}_{end1}), \xi_{1}(t)=1$, yielding that:\\

\begin{equation}
\xi_{r}(t)=\frac{1+\alpha_{l}\xi_{2}(t)}{1+\alpha_{l}\xi_{2}(t_{end1})},
\end{equation}
\begin{equation}
 [1+\alpha_{l}\xi_{2}(t_{end1})]\xi_{r}(t)=1+\alpha_{l}\xi_{2}(t).
\end{equation}

Besides, from Eq.(11) it can be obtained:\\

\begin{equation}
\xi_{r}(t)=\xi^{'}(t^{'})[1-\xi_{r}(t_{0})]+\xi_{r}(t_{0}).
\end{equation}

Substitution of $\xi_{r}(t)$ from Eq.(34) in Eq.(33) allows one obtaining:\\

\begin{equation}
\xi_{2}(t)=\xi^{'}_{2}(t^{'})=\frac{n_{2}}{\overline{n}}[\xi^{'}(t^{'})-1]+\xi_{2}(t_{end1}),
\end{equation}
\\where Eq.(27) has been taken into account. The fact is that $\xi_{2}(t_{end1})$ is unknown.
However if we set $t=t_{end2}$ in Eq.(35) we obtain:\\

\begin{equation}
\xi_{2}(t_{end1})=\frac{n_{2}}{\overline{n}}[1-\xi^{'}(t^{'}_{end2})]+1,
\end{equation}
\\so, finally, Eq.(35) can be written as:\\

\begin{equation}
\xi_{2}(t)=\xi^{'}_{2}(t^{'})=\frac{n_{2}}{\overline{n}}[\xi^{'}(t^{'})-\xi^{'}(t^{'}_{end2})]+1,
\end{equation}
\\this expression allows finding $\xi_{2}(t)$ (or $\xi^{'}_{2}(t^{'})$) by the knowledge of
$\xi^{'}(t^{'})$ from experiment. It is just our working
expression.

For computation of $\xi_{1}(t)\:\:(t_{0}<t<t_{end1})$, from Eq.(7) we have:\\

\begin{equation}
\xi_{1}(t)=\xi^{'}_{1}(t^{'})=\xi_{r}(t)[1+\alpha_{l}\xi_{2}(t_{end1})]-\alpha_{l}\xi_{2}(t).
\end{equation}

If we put Eq.(34) into Eq.(38) we obtain the working formula for this case:\\

\begin{equation}
\xi_{1}(t)=\xi^{'}_{1}(t^{'})=\xi^{'}(t^{'})[1-\xi_{r}(t_{0})][1+\alpha_{l}\xi_{2}(t_{end1})]+\xi_{r}(t_{0})[1+\alpha_{l}\xi_{2}(t_{end1})]-\alpha_{l}\xi_{2}(t).
\end{equation}

In what follows we shall study the two possible overlapping situations of the processes by the analysis of Eq.(39).\\

a)- Soft overlapping.

For times in the interval $(t_{0}<t_{a}\leq t\leq t_{b}<t_{end1})$
(or $0<t^{'}_{a}\leq t^{'}\leq t^{'}_{b}<t^{'}_{end1})$, and with
the help
of Eq.(24), the relationship (39) may be simplified:\\

\begin{equation}
\xi_{1}(t)=\xi^{'}_{1}(t^{'})=1+\frac{n_{1}}{n}[\xi^{'}(t^{'})-1],
\end{equation}
\\where we assumed that $\xi_{2}(t_{end1})-\xi_{2}(t)\approx0$.

b)-Hard overlapping.

For the time interval $(t_{0}<t_\alpha\leq t\leq
t_\beta<t_{end1})$, with the help of expression (30), the
relationship (39) can be written
now as:\\

\begin{equation}
\xi_{1}(t)=\xi^{'}_{1}(t^{'})=1+\frac{n_{1}}{n-\overline{n}}[\xi^{'}(t^{'})-1]+\frac{n_{1}\overline{n}}{n_{2}(n-\overline{n})}[\xi_{2}(t_{end1})-\xi_{2}(t)].
\end{equation}

We recall that, for a time interval $(t^{'}_{end1}<t^{'}_{\gamma}\leq t^{'}\leq t^{'}_{\delta}<t^{'}_{end2})$, the second process develops alone. Therefore, lineal independence of the exponential functions leads to $\overline{n} \:\:\mbox{and}\:\:n_{2}$ being well correlated. This can be shown by taking all the functional forms satisfying the exponential behaviour $\xi^{'}(t^{'})\:\:\mbox{vs}.\:\:\ln t^{'}$ in the interval $[t_{\gamma}, t_{\delta}]$, in the case of interest; $\alpha_{l}<0$, (fig. (1)), and satisfying, at the same time Eqs.(28) and (37) in the forms:\\

$(I)\:\:\:\frac{d\xi^{'}(t^{'})}{d\ln t^{'}}=\overline{n}Z(\xi^{'}), \:\:\mbox{with}\:\: \overline{n}<0$, \\
\\and\\

$(II)\:\:\:\xi^{'}(t^{'})=\frac{\overline{n}}{n_{2}}[\xi^{'}_{2}(t^{'})-1]+\xi^{'}(t^{'}_{end2})$.

The functions describing the experimental behavior $\xi^{'}(t^{'})\:\:\mbox{vs}.\:\:\ln t^{'}$ for $t^{'}\in[t^{'}_{\gamma}, t^{'}_{\delta}]$ are:\\

\begin{equation}
\xi^{'}(t^{'})=a-\exp
(-kt^{'})^{\overline{n}}\:\:\mbox{with}\:\:\overline{n}<0,
\end{equation}
\\and\\

\begin{equation}
\xi^{'}(t^{'})=b+\exp (-kt^{'})^{n}\:\:\mbox{with}\:\:n>0.
\end{equation}

Both satisfy expression (I). For the functional form (42) we have, (see Eq.(18)):\\

\begin{equation}
\frac{d\xi^{'}(t^{'})}{d\ln
t^{'}}=\overline{n}(kt^{'})^{\overline{n}}\exp(-kt^{'})^{\overline{n}}=\overline{n}Z(\xi^{'}),
\:\:\mbox{with}\:\: \overline{n}<0.
\end{equation}

For (43):\\

\begin{equation}
\frac{d\xi^{'}(t^{'})}{d\ln
t^{'}}=-n(kt^{'})^{n}\exp(-kt^{'})^{n}=-nZ(\xi^{'})=\overline{n}Z(\xi^{'}),
\:\:\mbox{with}\:\: \overline{n}=-n.
\end{equation}

It is worthwhile noting that only the functional form (43)
satisfies condition II with $n=n_{2}$. Therefore, since
$n=-\overline{n}$ this yields that $n_{2}=-\overline{n}$ .

Once the relation between $\overline{n}$  and $n_{2}$ is found, Eq.(41) simplifies:\\

\begin{equation}
\xi_{1}(t)=\xi^{'}_{1}(t^{'})=1+\frac{n_{1}}{n-\overline{n}}[\xi^{'}(t^{'})-\xi^{'}_{2}(t^{'}_{end1})-\exp(-k_{2}t^{'})^{n_{2}}].
\end{equation}

This equation allows obtaining the kinetic parameters $k_{1}$ and
$n_{1}$ of the first process in the case of hard overlapping of
processes for $t^{'}\in[t^{'}_{\alpha}, t^{'}_{\beta}]$.

With the further purpose of applying the kinetic theory of
overlapping processes, we compute   $\overline{n}$ as the slope of
the linear arrangement $\ln
\ln\frac{1}{1-\xi^{'}(t^{'})}\:\:\mbox{vs}\:\:\ln t^{'}$ for the
best fitting interval $t^{'}\in[t^{'}_{\gamma}, t^{'}_{\delta}]$
where $Z(\xi^{'})\approx Z(\xi_{2})$.

Functions $\xi^{'}_{2}(t^{'})$ are generated from Eq. (37) for the
same interval by means of an iterative procedure with an arbitrary
initial value $n_{2}$. A data $\xi^{'}_{2}(t^{'})$  is generated
for $t^{'}\in[t^{'}_{\gamma}, t^{'}_{\delta}]$ and by further
linear adjustment a new $n_{2}$ value is computed. The procedure
is repeated until a desired accuracy is reached. This way the
empiric parameters $n_{2}$ and $k_{2}$ are generated. Hence,
$\xi^{'}_{2}(t^{'}_{end1})$ can be evaluated following a JMA
expression and the kind of overlapping of processes can be
determined.

Once the kind of overlapping is known the correct expression (40)
or (46) is chosen, in order to obtain the kinetic parameters of
the first process.

A similar procedure enables one obtaining $n$ but taking the time
interval $t^{'}\in[t^{'}_{a}, t^{'}_{b}]$ or
$t^{'}\in[t^{'}_{\alpha}, t^{'}_{\beta}]$ according to the
overlapping type.

Once $n, \overline{n}, n_{2}, k_{2}$ and
$\xi^{'}_{2}(t^{'}_{end1})$ are known, a first data
$\xi^{'}_{1}(t^{'})$ may be computed according to whether kind of
overlapping Eqs. (40) or (46)
 one have for an arbitrary initial value $n_{1}$ in the above fitting interval
$t^{'}$. The data generated is linearly adjusted again to obtain a
new value $n_{1}$. The procedure is repeated until the desired
accuracy is reached while obtaining $n_{1}$ and $k_{1}$.\\

\section{Summary}

The set-up outlined in the present paper opens up new
possibilities for the study of isothermic phase transformations in
a variety of alloy systems. We require only the knowledge of just
some macroscopic parameter that is sensible to changes in the
fraction transformed. The kinetic parameters n and k can then be
computed following a very simple procedure. This allows, in turn,
obtaining a more complete information about the elementary
mechanisms of nucleation and growth of chemically different nuclei
(processes 1 and 2) that are competing for the same solute.

A key point of this set-up is that computing procedure works only
in the interval where the function $Z(\xi)$ is nearly constant,
leading to the experimental fraction transformed having behavior
nearly to JMA type expression.

In a forthcoming paper we apply the set-up presented here in the particular case of the
martensite tempering of a Mn low-alloy steel.\\

\begin{center}
{\bf Acknowledgements}
\end{center}
The authors wish to thank Dr. I. Quiros for reading the manuscript. Also wish to thank the Brazilian Foundations like CNPq and CAPES (EVM) for supporting part of this research.\\

\newpage

\begin{Large}

Figure Captions

\end{Large}

{\bf Figure 1}. Scheme of a dilatometric curve corresponding to
the isothermic tempering of a low-alloy steel.\\

{\bf Figure 2}. Functional dependence of $Z(\xi)$ vs. $\xi$.\\

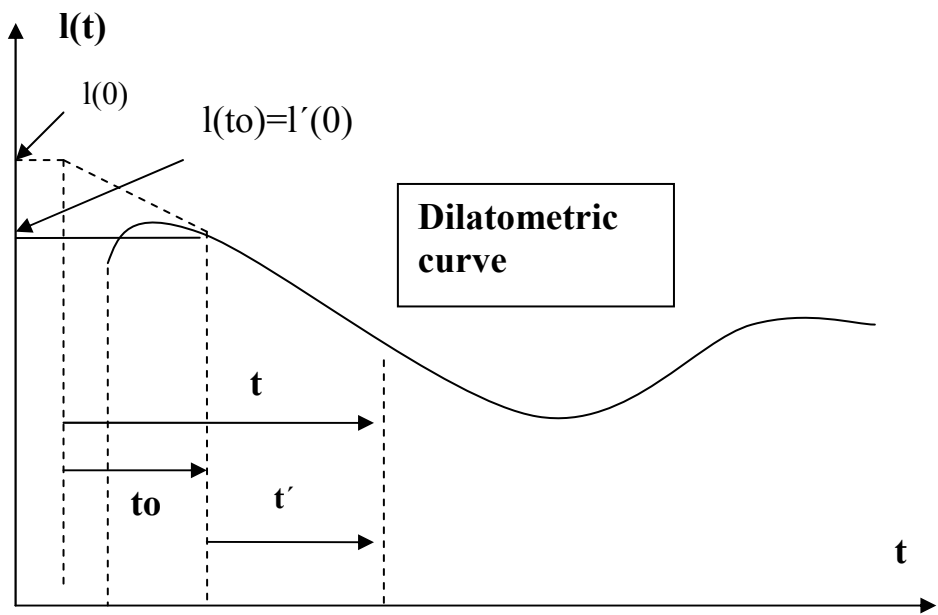
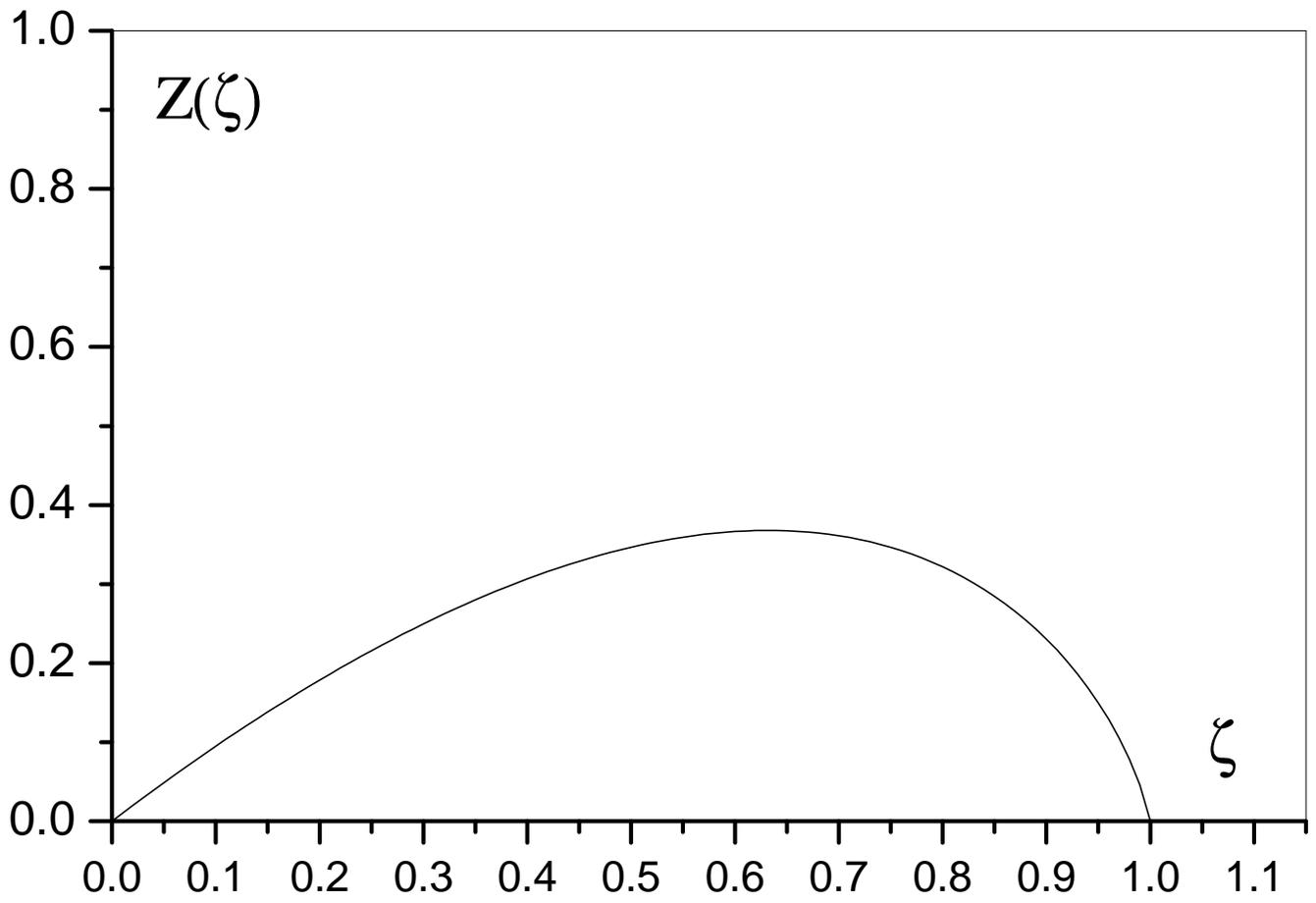
\end{document}